\documentclass[12pt,a4paper,english,nofootinbib]{revtex4}
\usepackage{lmodern}
\usepackage{lmodern}

\usepackage[T1]{fontenc}
\usepackage[latin9]{inputenc}
\usepackage{babel}
\usepackage{mathrsfs}
\usepackage{amsmath}
\usepackage{amssymb}
\usepackage{esint}
\usepackage[unicode=true,pdfusetitle,
 bookmarks=true,bookmarksnumbered=false,bookmarksopen=false,
 breaklinks=false,pdfborder={0 0 1},backref=false,colorlinks=false]
 {hyperref}

\makeatletter

\@ifundefined{textcolor}{}
{%
 \definecolor{BLACK}{gray}{0}
 \definecolor{WHITE}{gray}{1}
 \definecolor{RED}{rgb}{1,0,0}
 \definecolor{GREEN}{rgb}{0,1,0}
 \definecolor{BLUE}{rgb}{0,0,1}
 \definecolor{CYAN}{cmyk}{1,0,0,0}
 \definecolor{MAGENTA}{cmyk}{0,1,0,0}
 \definecolor{YELLOW}{cmyk}{0,0,1,0}
}

\usepackage{babel}
\usepackage{babel}
\usepackage{babel}
\usepackage{babel}
\usepackage{babel}
\usepackage{babel}
\usepackage{babel}
\usepackage{babel}
\usepackage{babel}
\usepackage{babel}
\usepackage{babel}
\usepackage{babel}
\usepackage{babel}
\usepackage{babel}
\usepackage{babel}
\usepackage{babel}
\usepackage{babel}
\usepackage{babel}
\usepackage{babel}
\usepackage{babel}

\usepackage{babel}
\usepackage{graphicx}
\def\b{\begin{equation}}
	\def\e{\end{equation}}

\@ifundefined{textcolor}{}{%
	\definecolor{BLACK}{gray}{0}
	\definecolor{WHITE}{gray}{1}
	\definecolor{RED}{rgb}{1,0,0}
	\definecolor{GREEN}{rgb}{0,1,0}
	\definecolor{BLUE}{rgb}{0,0,1}
	\definecolor{CYAN}{cmyk}{1,0,0,0}
	\definecolor{MAGENTA}{cmyk}{0,1,0,0}
	\definecolor{YELLOW}{cmyk}{0,0,1,0}
}

\usepackage{latexsym}\usepackage{bm}

\makeatother

\begin{document}
\title{{\normalsize{}{}{}{}{}{}{}{}{} EINSTEIN-YANG-MILLS THEORY:
GAUGE INVARIANT CHARGES AND LINEARIZATION INSTABILITY }}
\author{{\normalsize{}{}{}{}{}{}{}{}{}{}{}{}{}{}{}{}Emel Altas}}
\email{emelaltas@kmu.edu.tr}

\affiliation{Department of Physics, Karamanoglu Mehmetbey University, 70100, Karaman,
Turkey}
\author{{\normalsize{}{}{}{}{}{}{}{}{}{}{}{}{}{}{}{}Ercan Kilicarslan}}
\email{ercan.kilicarslan@usak.edu.tr}

\affiliation{Department of Mathematics, Usak University, 64200, Usak, Turkey}
\author{{\normalsize{}{}{}{}{}{}{}{}{}{}{}{}{}{}{}{}Bayram
Tekin}}
\email{btekin@metu.edu.tr}

\affiliation{Department of Physics, Middle East Technical University, 06800, Ankara,
Turkey}
\date{{\normalsize{}{}{}{}{}{}{}{}{}{{}{}{}{}\today}}}

\maketitle
We construct the gauge-invariant electric and magnetic charges in
Yang-Mills theory coupled to cosmological General Relativity (or any
other geometric gravity), extending the flat spacetime construction
of Abbott and Deser \citep{Abbott-Deser-nonabelian}. For non-vanishing
background gauge fields, the charges receive non-trivial contribution
from the gravity part. In addition, we study the constraints on the
first order perturbation theory and establish the conditions for linearization
instability: that is the validity of the first order perturbation
theory.

\section{{\normalsize{}{}{}{}{}{}{}{}{}{}{}{}{}{}{}{}INTRODUCTION}}

In 3+1 dimensions neither General Relativity \citep{Einstein-Pauli}
nor pure Yang-Mills theory \citep{Deser1976,Coleman} has solitonic
solutions. However, the coupled theory, the Einstein-Yang-Mills theory,
with or without a cosmological constant has various solitons. See
\citep{Bartnik-Mckinnon} for the first noted example in asymptotically
flat spacetimes, and \citep{Hosotani} for asymptotically anti-de
Sitter spacetimes. See \cite{Edery-Nakayama} for monopole type solutions
in $R^{2}$ gravity.

In this work we will work out the conserved charges of this coupled
system and also find the constraints in the linearization instability
of the first order perturbation theory. Conserved quantities in asymptotically
flat spacetimes for pure gravity was famously given in \citep{ADM},
which was generalized to asymptotically (anti) de Sitter spacetimes
in \citep{AD} and generalized to higher derivative gravity theories
in \citep{DT}. On the other hand, conserved gauge invariant charges
in pure Yang-Mills theory was constructed by Abbott and Deser \citep{Abbott-Deser-nonabelian}.

Here in the first part of this work we follow the Abbott-Deser construction
for a dynamical curved background with generically an asymptotically
(A)dS behavior.

The second problem we study is the question of the validity of the
perturbation theory in the Einstein-Yang-Mills system. It is well
known \citep{Fischer-Marsden,Deser-Brill,Deser-Bruhat,Moncrief,Arms-Marsden,Fischer-Marsden-Moncrief,Marsden,Girbau-Bruna,emelchiral,emeltez,altasuzunmakale}
that not all perturbative solutions come from the linearization of
a possible exact solution. If that happens, one speaks of linearization
instability and the perturbation thus fails. To have a linearization
stable theory the first order perturbative solution must satisfy an
integral constraint. We shall find this for the Einstein-Yang-Mills
theory.

Before we study the Einstein-Yang-Mills system in full detail, let
us give our conventions \citep{Weinberg} and recap the flat space
construction \citep{Abbott-Deser-nonabelian}. We will work in $D=3+1$
dimensions exclusively, but the discussion can be extended to other
dimensions with the caveat that both pure Yang-Mills theory and pure
General Relativity might have solitonic solutions for $D>3+1$. We
use the mostly plus signature $(-+++)$ and assume a compact Lie group
${\mathsf{G}}$ with the Lie algebra $\ensuremath{{\cal {G}}}$ given
as 
\begin{equation}
\left[T_{a},T_{b}\right]=iC_{abc}T^{c},
\end{equation}
with $C_{abc}$ real. In the adjoint representation we write $(T_{a}^{Ad})^{b}\thinspace_{c}:=-iC^{b}\thinspace_{ca}$;
and defining $(D_{\mu}\psi)_{n}:=\partial_{\mu}\psi_{n}-iA_{\mu}^{a}(T_{a})_{n}^{k}\psi_{k}$
in flat spacetime we have 
\begin{equation}
\mathscr{L}=-\frac{1}{4}F_{\mu\nu}^{a}F_{a}^{\mu\nu}+\text{\ensuremath{\mathscr{L}}}_{matter}(\psi,D_{\mu}\psi),
\end{equation}
with the field equations 
\begin{equation}
\partial_{\mu}F_{a}^{\mu\nu}=-J_{a}^{\nu}.
\end{equation}
The current 
\begin{equation}
J_{a}^{\mu}=-F_{c}^{\mu\nu}C_{cab}A_{b\nu}-i\frac{\partial\text{\ensuremath{\mathscr{L}}}_{matter}}{\partial(D_{\mu}\psi)_{n}}(T_{a})_{n}^{k}\psi_{k},
\end{equation}
is partially conserved 
\begin{equation}
\partial_{\mu}J_{a}^{\mu}=0,
\end{equation}
and hence yields the conserved charges 
\begin{equation}
Q_{a}:=\int J_{a}^{0}d^{3}x.
\end{equation}
But these charges are gauge-covariant, not gauge-invariant. To get
gauge--invariant charges, one can employ the AD technique \citep{AD}
which is based on the following observation. Assuming the Yang-Mills
coupling $g_{YM}=1$, without loss of generality, we can define the
matrix valued gauge field and the field strength 
\begin{equation}
\hat{A}_{\mu}:=T^{a}A_{\mu}^{a},~~~~~~~~~~~~~~\hat{F}_{\mu\nu}:=T^{a}F_{\mu\nu}^{a}.
\end{equation}
Let the unitary matrix $\hat{U}$ be in the same representation as
$T^{a}$, then the gauge transformed gauge field reads 
\begin{equation}
\hat{A}_{\mu}^{\hat{U}}=\hat{A}_{\mu}+\hat{U}^{-1}D_{\mu}\hat{U},
\end{equation}
with the gauge-covariant derivative defined as 
\begin{equation}
D_{\mu}\hat{U}:=\nabla_{\mu}\hat{U}+[\hat{A}_{\mu},\hat{U}],
\end{equation}
and the field strength transforms as usual 
\begin{equation}
\hat{F}_{\mu\nu}^{\hat{U}}=\hat{U}^{-1}\hat{F}_{\mu\nu}\hat{U}.
\end{equation}
Under an infinitesimal transformation $\hat{U}\cong1+\hat{\xi}$,
one gets 
\begin{equation}
\delta\hat{A}_{\mu}=D_{\mu}\hat{\xi},~~~~~~~~~~~~~~\delta\hat{F}_{\mu\nu}=[\hat{F}_{\mu\nu},\hat{\xi}].
\end{equation}
So clearly, $D_{\mu}\hat{\xi}=0$ defines the symmetries of a ``background''
field $\hat{A}_{\mu}$ which we shall denote as $\bar{\hat{A}}_{\mu}$
from now on; clearly $[\hat{F}_{\mu\nu},\hat{\xi}]=0$. In the space
of gauge fields, $\hat{\xi}$ acts like a Killing vector akin to the
spacetime Killing vectors $\delta g_{\mu\nu}=\nabla_{\mu}X_{\nu}+\nabla_{\nu}X_{\mu}=0$.
As there can be more than one solution to $D_{\mu}\hat{\xi}=0$, we
shall put an index to denote the elements of the symmetry set and
write as $\hat{\xi}^{s}$, which is exactly the correct matrix that
will turn $J_{a}^{\mu}$ to be a gauge invariant current since $\text{Tr}(\hat{\xi}\hat{J}_{\mu})$
is gauge-invariant for $\hat{J}_{\mu}=J_{\mu}^{a}T^{a}$. But this
procedure requires a choice of background gauge-field and hence it
must be done carefully. Instead of repeating the full details of the
flat space construction, we now study the curved space version which
also includes the flat space as a special case.

\section{Construction of the conserved charges in the Einstein-Yang-Mills
system}

The following construction works for any gravity theory based on Riemannian
geometry with a Lagrangian of the generic form, but for the sake of
concreteness, we shall take the gravity sector to be given as the
Einstein-Hilbert Lagrangian. The coupled action reads 
\begin{equation}
S=\intop_{{\mathcal{M}}}d^{4}x\sqrt{-g}\thinspace\Bigg(\frac{R-2\Lambda}{2\kappa}-\frac{1}{4}F_{\mu\nu}^{a}F_{a}^{\mu\nu}+\text{\ensuremath{\mathscr{L}}}_{matter}\Bigg).
\end{equation}
As long as the Yang-Mills and the matter fields decay sufficiently
fast at spatial infinity, the conserved energy, momentum, angular
momentum as constructed, say in \citep{Abbott-Deser-nonabelian,DT}
are intact, so we will not repeat these well-established discussions
here, but work out the Yang-Mills part in some detail. Variation with
respect to $\hat{A}_{\mu}$ yields 
\begin{equation}
D_{\mu}\hat{F}^{\mu\nu}=\nabla_{\mu}\hat{F}^{\mu\nu}+[\hat{A}_{\mu},\hat{F}^{\mu\nu}]=\hat{J}^{\nu},\label{field equations1}
\end{equation}
where 
\begin{equation}
\hat{F}^{\mu\nu}:=\nabla^{\mu}\hat{A}^{\nu}-\nabla^{\nu}\hat{A}^{\mu}+[\hat{A}^{\mu},\hat{A}^{\nu}].\label{field strength}
\end{equation}
The field strength satisfies the Bianchi identity 
\begin{equation}
D_{\alpha}\hat{F}_{\mu\nu}+D_{\mu}\hat{F}_{\nu\alpha}+D_{\nu}\hat{F}_{\alpha\mu}=0,
\end{equation}
and with the normalization $\text{Tr}(T^{a}T^{b})=\frac{1}{2}\delta^{ab}$,
one has in components 
\begin{equation}
F_{\mu\nu}^{a}=\partial_{\mu}A_{\nu}^{a}-\partial_{\nu}A_{\mu}^{a}+C^{a}\thinspace_{bc}A_{\mu}^{b}A_{\nu}^{c}.
\end{equation}
Using (\ref{field equations1}) we will construct the gauge-invariant
electric field, while the magnetic charge will follow from the Bianchi
identity. Assume now that for $\hat{J}^{\nu}=0$, the background matrix
$\bar{\hat{A}}_{\mu}$ solves the source-free equation $\bar{D}_{\mu}\bar{\hat{F}}^{\mu\nu}=0$;
and we expand the field equations about this solution as~\footnote{Please see Appendix-A for an extended discussion of the expansion
of the field equations up to and including second order in perturbation
theory.} 
\begin{equation}
\hat{A}_{\mu}=\bar{\hat{A}}_{\mu}+\lambda\hat{a}_{\mu}+\frac{\lambda^{2}}{2}\hat{b}_{\mu}+{\mathcal{O}}(\lambda^{3}),\label{expansionofgaugefiedl}
\end{equation}
where $\lambda$ is a small parameter. As we are in a dynamical background
spacetime, the metric also receives perturbations which we shall write
as 
\begin{equation}
g_{\mu\nu}=\bar{g}_{\mu\nu}+\tau h_{\mu\nu}+\frac{\tau^{2}}{2}k_{\mu\nu}+{\mathcal{O}}(\tau^{3}),\label{expansionofmetric}
\end{equation}
with $\tau$ being a different small parameter. Under these expansions,
the full equation split as 
\begin{equation}
D_{\mu}\hat{F}^{\mu\nu}=\bar{D}_{\mu}\bar{\hat{F}}^{\mu\nu}+(D_{\mu}\hat{F}^{\mu\nu})^{(1)}+(D_{\mu}\hat{F}^{\mu\nu})^{(2)}+...=J^{\nu},\label{denkk}
\end{equation}
and by assumption the zeroth order term vanishes in the absence of
a source 
\begin{equation}
\bar{D}_{\mu}\bar{\hat{F}}^{\mu\nu}=\bar{\nabla}_{\mu}\bar{\hat{F}}^{\mu\nu}+[\bar{\hat{A}}_{\mu},\bar{\hat{F}}^{\mu\nu}]=0.\label{background field equations}
\end{equation}
At the linear order one finds 
\begin{equation}
(D_{\mu}\hat{F}^{\mu\nu})^{(1)}=\bar{D}_{\mu}\left(\lambda(\bar{D}^{\mu}\hat{a}^{\nu}-\bar{D}^{\nu}\hat{a}^{\mu})+\tau(\bar{\hat{F}}^{\sigma\mu}h_{\sigma}^{\nu}-\bar{\hat{F}}^{\sigma\nu}h_{\sigma}^{\mu}+\frac{1}{2}\bar{\hat{F}}^{\mu\nu}h)\right)+\lambda[\hat{a}_{\mu},\bar{\hat{F}}^{\mu\nu}].\label{D_muF^munu-firstorder}
\end{equation}
Similarly, the second order expansion, $(D_{\mu}\hat{F}^{\mu\nu})^{(2)}$,
reads 
\begin{eqnarray}
 &  & (D_{\mu}\hat{F}^{\mu\nu})^{(2)}=\bar{D}_{\mu}\left((\hat{F}^{\mu\nu})^{(2)}+\frac{\tau}{2}(\hat{F}^{\mu\nu})^{(1)}h+\frac{\tau^{2}}{4}\bar{\hat{F}}^{\mu\nu}(k-h_{\rho\sigma}h^{\rho\sigma})\right)\nonumber \\
 &  & ~~~~~~~~~~~~~~~~~~-\frac{\tau}{2}h\bar{D}_{\mu}(\hat{F}^{\mu\nu})^{(1)}+\lambda[\hat{a}_{\mu},(\hat{F}^{\mu\nu})^{(1)}]+\frac{\lambda^{2}}{2}[\hat{b}_{\mu},\bar{\hat{F}}^{\mu\nu}].\label{secorder}
\end{eqnarray}
Moving all the higher order terms to the right-hand side, we can recast
(\ref{denkk}) as 
\begin{equation}
(D_{\mu}\hat{F}^{\mu\nu})^{(1)}=\hat{{\mathcal{J}}}^{\nu},\label{lindenk}
\end{equation}
where the current 
\begin{equation}
\hat{{\mathcal{J}}}^{\nu}:=\hat{J}^{\nu}-(D_{\mu}\hat{F}^{\mu\nu})^{(2)}-...
\end{equation}
is composed of the matter current as well as all the terms beyond
the linear one coming from the expansion. The crucial point is that
this current is covariantly conserved with respect to the background
connection explicitly, $\bar{D}_{\nu}\hat{{\mathcal{J}}}^{\nu}=0$.\footnote{To see the direct computation for the conservation of the current
see Appendix-B.}

Finally, substituting (\ref{D_muF^munu-firstorder}) in (\ref{lindenk}),
one finds 
\begin{equation}
\bar{D}_{\mu}\left(\lambda(\bar{D}^{\mu}\hat{a}^{\nu}-\bar{D}^{\nu}\hat{a}^{\mu})+\tau(\bar{\hat{F}}^{\sigma\mu}h_{\sigma}^{\nu}-\bar{\hat{F}}^{\sigma\nu}h_{\sigma}^{\mu}+\frac{1}{2}\bar{\hat{F}}^{\mu\nu}h)\right)+\lambda[\hat{a}_{\mu},\bar{\hat{F}}^{\mu\nu}]=\hat{{\mathcal{J}}}^{\nu}.\label{linearizedfield equations}
\end{equation}
Covariantly conserved current does not immediately yield a conserved
charge; to get a partially conserved current, we appeal to the symmetries
of the background gauge field as discussed in the previous section.
So we assume the existence of some (but at least one) background gauge
covariant matrices $\bar{\hat{\xi}}^{s}$ such that 
\begin{equation}
\bar{D}_{\mu}\bar{\hat{\xi}}^{s}=\bar{\nabla}_{\mu}\bar{\hat{\xi}}^{s}+[\bar{\hat{A}}_{\mu},\bar{\hat{\xi}}^{s}]=0,
\end{equation}
which yields $\left[\bar{D}_{\nu},\bar{D}_{\mu}\right]\bar{\hat{\xi}}^{s}=0=[\bar{\hat{F}}_{\nu\mu},\bar{\hat{\xi}}^{s}].$
Since $\bar{D}_{\nu}\hat{{\mathcal{J}}}^{\nu}=0$ and $\bar{D}_{\nu}\bar{\hat{\xi}}^{s}=0$,
we can write 
\begin{equation}
\sqrt{-\bar{g}}\bar{D}_{\nu}\text{Tr}(\bar{\hat{\xi}}^{s}\hat{{\mathcal{J}}}^{\nu})=\sqrt{-\bar{g}}\bar{\nabla}_{\nu}\text{Tr}(\bar{\hat{\xi}}^{s}\hat{{\mathcal{J}}}^{\nu})=\partial_{\nu}\left(\sqrt{-\bar{g}}\text{Tr}(\bar{\hat{\xi}}^{s}\hat{{\mathcal{J}}}^{\nu})\right)=0,
\end{equation}
which can be used to express the conserved electric charges\footnote{For the details of the calculation see Appendix-C.}
for each background gauge symmetry as: 
\begin{equation}
Q_{\text{E}}^{s}:=\frac{1}{4\pi}\intop_{\Sigma}d^{3}x~\sqrt{\bar{\gamma}}\text{Tr}(\bar{\hat{\xi}}^{s}\hat{{\mathcal{J}}}^{0}),\label{conservedcharges1}
\end{equation}
where we assumed that the four dimensional spacetime ${\mathcal{M}}$
is diffeomorphic to $\Sigma\times\mathbb{R}$ and $\bar{\gamma}$
denotes the induced metric on the spatial hypersurface. Using the
explicit form of the current and employing the Stokes' theorem, one
arrives at 
\begin{equation}
Q_{\text{E}}^{s}=\frac{1}{4\pi}\intop_{\partial\Sigma}d^{2}x~\sqrt{\bar{\beta}}\sigma_{i}\text{Tr}\left(\bar{\hat{\xi}}^{s}\Bigl(\lambda(\bar{D}^{i}\hat{a}^{0}-\bar{D}^{0}\hat{a}^{i})+\tau(\bar{\hat{F}}^{0i}h_{0}^{0}+\bar{\hat{F}}^{ki}h_{k}^{0}+\bar{\hat{F}}^{0k}h_{k}^{i}+\frac{h}{2}\bar{\hat{F}}^{i0})\Bigr)\right),\label{guzelsonuc}
\end{equation}
where $\bar{\beta}$ is the two dimensional induced metric on the
boundary of the hypersurface and $\sigma_{i}$ is its unit one form.
Observe that if the background gauge field is chosen to be pure gauge
or zero, then the order $\tau$ term in the charge expression vanishes
and the gauge-invariant electric charges have the same form as their
flat spacetime versions \citep{Abbott-Deser-nonabelian}, while generically
gravity contributes in a nontrivial way.

Magnetic charge discussion follows similarly but now one employs the
Bianchi identity which can be written with the help of the dual of
the field strength as 
\begin{equation}
D_{\mu}\thinspace^{\star}\hat{F}^{\mu\nu}=0,
\end{equation}
where $^{\star}\hat{F}^{\mu\nu}:=\frac{1}{2\sqrt{-g}}\epsilon^{\mu\nu\rho\sigma}\hat{F}_{\rho\sigma}$.
More explicitly the identity can be written as 
\begin{equation}
\frac{1}{2\sqrt{-g}}\epsilon^{\mu\nu\rho\sigma}\left(\partial_{\mu}\hat{F}_{\rho\sigma}+[\hat{A}_{\mu},\hat{F}_{\rho\sigma}]\right)=0,
\end{equation}
which is the same as the expression in the flat spacetime case. So,
expanding the gauge field about a background $\bar{\hat{A}}_{\mu}$,
and the metric tensor about $\bar{g}_{\mu\nu}$ one arrives at 
\begin{equation}
^{\bar{\star}}\hat{\mathcal{J}}^{\nu}=\bar{D}_{\mu}\thinspace^{\star}(\hat{F}^{\mu\nu})^{\left(1\right)}+\lambda[\hat{a}_{\mu},{}^{\bar{\star}}\bar{\hat{F}}^{\mu\nu}],
\end{equation}
with the linear part of the dual field strength given as $^{\star}(\hat{F}^{\mu\nu})^{(1)}=\frac{1}{2\sqrt{-\bar{g}}}\bar{\epsilon}^{\mu\nu\rho\sigma}\lambda(\bar{D}_{\rho}\hat{a}_{\sigma}-\bar{D}_{\sigma}\hat{a}_{\rho})$
and the background dual field as $^{\bar{\star}}\bar{\hat{F}}^{\mu\nu}=\frac{1}{2\sqrt{-\bar{g}}}\bar{\epsilon}^{\mu\nu\rho\sigma}(\bar{D}_{\rho}\bar{\hat{A}}_{\sigma}-\bar{D}_{\sigma}\bar{\hat{A}}_{\rho})$.
Then the conserved magnetic charges can be written as 
\begin{equation}
Q_{M}^{s}=\frac{1}{4\pi}\intop_{\partial\Sigma}d^{2}x\,\sigma_{i}\thinspace\text{Tr}\left(\bar{\xi}^{s}\thinspace^{\bar{\star}}(F^{i0})^{\left(1\right)}\right),\label{eq:33}
\end{equation}
which has the same form as its flat space version \citep{Abbott-Deser-nonabelian}.
The magnetic charges are topological: as can be seen from comparing
equations (29) and (33) the metric $\sqrt{\bar{\beta}}$ does not
explicitly appear in (33). Instead the Hodge dual appears which just
is used to define the magnetic field. Hence we can equivalently express
(\ref{eq:33}) as follows
\[
Q_{M}=\frac{1}{4\pi}\intop_{\partial\varSigma}d^{2}x\thinspace\sigma_{i}\thinspace B^{(s)i},
\]
where 
\[
B^{(s)i}=Tr(\bar{\xi}^{(s)}\thinspace{}^{\bar{*}}(F^{i0})^{(1)}).
\]

\section{Linearization Instability}

In nonlinear theories, there are some cases for which the first order
perturbation theory is constrained at the second order. When this
happens, one speaks of the theory having a linearization instability
about the zeroth order (or the background solution). This topic is
rather extensive: see \citep{Fischer-Marsden,Deser-Brill,Deser-Bruhat,Moncrief,Arms-Marsden,Marsden,Girbau-Bruna,altasuzunmakale,emelchiral};
and for a relevant review of the literature we would like to refer
the reader to the recent PhD thesis \citep{emeltez}, where the issue
is elaborated in sufficient detail. Here let us study the linearization
instability issue in the Gravity-Yang-Mills system. {[}Einstein gravity
can be taken as a concrete example, but generic gravity theories can
exhibit nontrivial linearization instability behavior as discussed
in \citep{altasuzunmakale,emelchiral}.{]} We first assume a spacetime
with noncompact hypersurfaces and at the end concentrate on the case
of compact hypersurfaces without a boundary. Let us go back to the
Yang-Mills equation with $J^{\nu}=0$, expand again up to second order
in the gauge-field and the metric perturbation to get 
\begin{equation}
(\bar{D}_{\mu}\bar{\hat{F}}^{\mu\nu})\cdot(\bar{\hat{A}},\bar{g})+\left(D_{\mu}F^{\mu\nu}\right)^{(1)}\cdot(\hat{a},h)+\left(D_{\mu}F^{\mu\nu}\right)^{(1)}\cdot(\hat{b},k)+\left(D_{\mu}F^{\mu\nu}\right)^{(2)}\cdot(\hat{a}^{2},h^{2},\hat{a}h)+...=0,\label{expansion}
\end{equation}
where the center dot notation means, for example, $\left(D_{\mu}F^{\mu\nu}\right)^{(1)}$
operator is evaluated at the first order expansion of the gauge field
and the metric tensor $(\hat{a},h)$. By assumption, we have $(\bar{D}_{\mu}\bar{F}^{\mu\nu})\cdot(\bar{\hat{A}},\bar{g})=0$,
which together with the gravity sector, determine the background solutions
$(\bar{A}_{\mu},\bar{g}_{\mu\nu})$ up to gauge degrees of freedom,
of course.

Similarly, by assumption, we have $\left(D_{\mu}F^{\mu\nu}\right)^{(1)}\cdot(\hat{a},h)=0$,
which together with the linearized part of the gravity sector, determine
the linearized solutions $(\hat{a}_{\mu},h_{\mu\nu})$, again up to
gauge transformations. So the second order terms are determined from
the equation 
\begin{equation}
\left(D_{\mu}F^{\mu\nu}\right)^{(1)}\cdot(\hat{b},k)+\left(D_{\mu}F^{\mu\nu}\right)^{(2)}\cdot(\hat{a}^{2},h^{2},\hat{a}h)=0,\label{sart}
\end{equation}
which basically says that once $(\hat{a}_{\mu},h_{\mu\nu})$ are found
from the linearized equations, $-\left(D_{\mu}F^{\mu\nu}\right)^{(2)}\cdot(\hat{a}^{2},h^{2},\hat{a}h)$
acts like a source term for the second order perturbations $(\hat{b}_{\mu},k_{\mu\nu})$.
If this happens then the first order perturbation theory is intact
and improvable and moreover, linearized \textit{{solutions}} obtained
from the linearized \textit{{equations}} can come from the linearization
of some {\textit{exact solutions}}. Please see the diagram in \citep{altasuzunmakale}
that depicts this commumativity.

So the necessary and sufficient condition for linearization stability
is that (\ref{sart}) should not constrain the first order solutions
$(\hat{a}_{\mu},h_{\mu\nu})$ and it should determine the second order
solutions $(\hat{b}_{\mu},k_{\mu\nu})$ up to gauge transformations.
But clearly this is very hard to check for all linear solutions of
the theory, so in what follows let us find a weaker (necessary) condition.
This condition will be in the form of an integral whose purely gravitational
analog is called the Taub charge \citep{Taub} and see the following
recent discussion \citep{emelson}. From (\ref{sart}), we have 
\begin{equation}
\intop_{\Sigma}d^{3}x~\sqrt{\bar{\gamma}}\,\text{Tr}\left(\bar{\hat{\xi}}^{s}(D_{\mu}\hat{F}^{\mu0})^{(1)}\cdot(\hat{b},k)+\bar{\hat{\xi}}^{s}(D_{\mu}\hat{F}^{\mu0})^{(2)}\cdot(\hat{a}^{2},h^{2},\hat{a}h)\right)=0.\label{sart2}
\end{equation}
The first term in the integrand is of the same form as the first order
term $\left(D_{\mu}F^{\mu0}\right)^{(1)}\cdot(\hat{a},h)$, albeit
now evaluated at the second order fields instead of the first order
ones. So, obviously this piece can be written as a boundary term as
(\ref{guzelsonuc}) with the substitution $(\hat{a},h)\rightarrow(\hat{b},k)$.

The second term in the integrand requires more work, it is not clear
at all if it can be written as a boundary integral. Nevertheless,
to write some parts of $(D_{\mu}\hat{F}^{\mu0})^{(2)}\cdot(\hat{a}^{2},h^{2},\hat{a}h)$
as a boundary term, we use the explicit form of the second order expansion
(\ref{secorder}): 
\begin{eqnarray}
(D_{\mu}\hat{F}^{\mu\nu})^{(2)}\cdot(\hat{a}^{2},h^{2},\hat{a}h) & = & \bar{D}_{\mu}\left((\hat{F}^{\mu\nu})^{(2)}+\frac{\tau}{2}(\hat{F}^{\mu\nu})^{(1)}h+\frac{\tau^{2}}{4}\bar{\hat{F}}^{\mu\nu}(k-h_{\rho\sigma}h^{\rho\sigma})\right)\nonumber \\
 &  & -\frac{\tau}{2}h\bar{D}_{\mu}(\hat{F}^{\mu\nu})^{(1)}+\lambda[\hat{a}_{\mu},(\hat{F}^{\mu\nu})^{(1)}]+\frac{\lambda^{2}}{2}[\hat{b}_{\mu},\bar{\hat{F}}^{\mu\nu}],\label{uzundenk}
\end{eqnarray}
where 
\begin{eqnarray}
 &  & (\hat{F}^{\mu\nu})^{(2)}=\frac{\lambda^{2}}{2}\left(\bar{D}^{\mu}\hat{b}^{\nu}-\bar{D}^{\nu}\hat{b}^{\mu}+2\left[\hat{a}^{\mu},\hat{a}^{\nu}\right]\right)+\tau\lambda\left(h^{\nu\sigma}(\bar{D}_{\sigma}\hat{a}^{\mu}-\bar{D}^{\mu}\hat{a}_{\sigma})+h^{\mu\sigma}(\bar{D}^{\nu}\hat{a}_{\sigma}-\bar{D}_{\sigma}\hat{a}^{\nu})\right)\nonumber \\
 &  & ~~~~~~~~~~~+\frac{\tau^{2}}{2}\left(\bar{\hat{F}}^{\mu\sigma}(2h^{\nu\lambda}h_{\lambda\sigma}-k_{\sigma}^{\nu})-\bar{\hat{F}}^{\nu\sigma}(2h^{\mu\lambda}h_{\lambda\sigma}-k_{\sigma}^{\mu})+2\bar{\hat{F}}_{\sigma\rho}h^{\mu\sigma}h^{\nu\rho}\right).
\end{eqnarray}
Inserting this expression into (\ref{uzundenk}) obtains 
\begin{equation}
\left(D_{\mu}F^{\mu\nu}\right)^{(2)}\cdot(\hat{a}^{2},h^{2},\hat{a}h)=\bar{D}_{\mu}\hat{\mathcal{X}}^{\mu\nu}-\frac{\tau}{2}h\bar{D}_{\mu}(\hat{F}^{\mu\nu})^{(1)}+\lambda[\hat{a}_{\mu},(\hat{F}^{\mu\nu})^{(1)}],\label{secondordereq}
\end{equation}
where we have introduced an antisymmetric field, $\hat{\mathcal{X}}^{\mu\nu}$,
to express the result in a more compact form. Direct calculation yields
\begin{align}
\hat{\mathcal{X}}^{\mu\nu} & =\frac{\tau}{2}(\hat{F}^{\mu\nu})^{(1)}h+\lambda^{2}\left[\hat{a}^{\mu},\hat{a}^{\nu}\right]-\frac{\tau^{2}}{4}\bar{\hat{F}}^{\mu\nu}h_{\rho\sigma}h^{\rho\sigma}+\tau^{2}\left(\bar{\hat{F}}^{\mu\sigma}h^{\nu\lambda}h_{\lambda\sigma}-\bar{\hat{F}}^{\nu\sigma}h^{\mu\lambda}h_{\lambda\sigma}+\bar{\hat{F}}_{\sigma\rho}h^{\mu\sigma}h^{\nu\rho}\right)\nonumber \\
 & \ ~~\ ~~\ ~\ ~~\ ~+\tau\lambda\left(h^{\nu\sigma}(\bar{D}_{\sigma}\hat{a}^{\mu}-\bar{D}^{\mu}\hat{a}_{\sigma})+h^{\mu\sigma}(\bar{D}^{\nu}\hat{a}_{\sigma}-\bar{D}_{\sigma}\hat{a}^{\nu})\right).\label{xmunu}
\end{align}
Then from (\ref{secondordereq}), one finds 
\begin{eqnarray}
 &  & \text{Tr}\left(\bar{\hat{\xi}}^{s}(D_{\mu}\hat{F}^{\mu0})^{(2)}\cdot(\hat{a}^{2},h^{2},\hat{a}h)\right)=\bar{\nabla}_{i}\text{Tr}\left(\bar{\xi}^{s}\hat{\mathcal{X}}^{i0}\right)-\frac{\tau}{2}\text{Tr}\left(\bar{\hat{\xi}}^{s}h\bar{D}_{i}(\hat{F}^{i0})^{(1)}\right)\nonumber \\
 &  & ~~\ ~~\ ~\ ~~\ ~~\ ~~\ ~\ ~~\ ~~\ ~~\ ~\ ~~\ ~~\ ~~\ ~\ ~~\ +\lambda\text{Tr}\left(\bar{\hat{\xi}}^{s}[\hat{a}_{i},(\hat{F}^{i0})^{(1)}]\right).
\end{eqnarray}
Since $\text{Tr}\bigl(\bar{\hat{\xi}}^{s}[\hat{a}_{i},(\hat{F}^{i0})^{(1)}]\bigr)=\text{Tr}\bigl([(\hat{F}^{i0})^{(1)},\bar{\hat{\xi}}^{s}]\hat{a}_{i}\bigr)$,
we have 
\begin{align}
 & \intop_{\Sigma}d^{3}x~\sqrt{\bar{\gamma}}\text{Tr}\left(\bar{\hat{\xi}}^{s}(D_{\mu}\hat{F}^{\mu0})^{(2)}\cdot(\hat{a}^{2},h^{2},\hat{a}h)\right)=\intop_{\Sigma}d^{3}x~\partial_{i}\left(\sqrt{\bar{\gamma}}\text{Tr}\bigl(\bar{\hat{\xi}}^{s}\hat{\mathcal{X}}^{i0}\bigr)\right)\nonumber \\
 & \ ~~\ ~~\ ~~\ ~\ ~+\lambda\intop_{\Sigma}d^{3}x~\sqrt{\bar{\gamma}}\text{Tr}\left([(\hat{F}^{i0})^{(1)},\bar{\hat{\xi}}^{s}]\hat{a}_{i}\right)-\frac{\tau}{2}\intop_{\Sigma}d^{3}x~\sqrt{\bar{\gamma}}\text{Tr}\left(\bar{\hat{\xi}}^{s}h\bar{D}_{i}(\hat{F}^{i0})^{(1)}\right).\label{denk}
\end{align}
If we use the first order equation 
\begin{equation}
\bar{D}_{i}(\hat{F}^{i0})^{(1)}=\frac{\tau}{2}\bar{D}_{i}\bigl(h\bar{\hat{F}}^{i0}\bigr)+\lambda[\hat{a}_{i},\bar{\hat{F}}^{i0}],
\end{equation}
(\ref{denk}) reduces to 
\begin{align}
 & \intop_{\Sigma}d^{3}x~\sqrt{\bar{\gamma}}\text{Tr}\left(\bar{\hat{\xi}}^{s}(D_{\mu}\hat{F}^{\mu0})^{(2)}\cdot(\hat{a}^{2},h^{2},\hat{a}h)\right)=\intop_{\Sigma}d^{3}x~\partial_{i}\left(\sqrt{\bar{\gamma}}\text{Tr}\bigl(\bar{\hat{\xi}}^{s}\hat{\mathcal{X}}^{i0}\bigr)\right)\nonumber \\
 & \ ~~\ ~~\ ~~\ ~\ ~+\lambda\intop_{\Sigma}d^{3}x~\sqrt{\bar{\gamma}}\text{Tr}\left([(\hat{F}^{i0})^{(1)},\bar{\hat{\xi}}^{s}]a_{i}\right)-\frac{\tau^{2}}{8}\intop_{\Sigma}d^{3}x~\partial_{i}\left(\sqrt{\hat{\gamma}}\text{Tr}\bigl(\bar{\hat{\xi}}^{s}h^{2}\bar{\hat{F}}^{i0}\bigr)\right).\label{denk-1}
\end{align}
So from (\ref{sart2}) we arrive at 
\begin{equation}
\lambda\intop_{\Sigma}d^{3}x~\sqrt{\bar{\gamma}}\text{Tr}\left([(\hat{F}^{i0})^{(1)},\bar{\hat{\xi}}^{s}]a_{i}\right)=\int_{\partial\Sigma}d^{2}x~\sqrt{\bar{\beta}}{\mathcal{I}},\label{sart3}
\end{equation}
where we know ${\mathcal{I}}$ from (\ref{sart2}) and (\ref{denk-1})
explicitly so we need not depict it again. Consider now the case for
which all the fields decay sufficiently fast, such that the boundary
term on the right-hand side vanishes, or the case when the hyperspace
is compact without a boundary ($\partial\Sigma=0$), then we get an
integral constraint in the bulk for the linearized solutions: 
\begin{equation}
\intop_{\Sigma}d^{3}x~\sqrt{\bar{\gamma}}\text{Tr}\left([(\hat{F}^{i0})^{(1)},\bar{\hat{\xi}}^{s}]\hat{a}_{i}\right)=0,
\end{equation}
which reads explicitly as 
\begin{equation}
\intop_{\Sigma}d^{3}x~\sqrt{\bar{\gamma}}\text{Tr}\left([\bar{D}^{i}\hat{a}^{0}-\bar{D}^{0}\hat{a}^{i},\bar{\hat{\xi}}^{s}]\hat{a}_{i}\right)=0.
\end{equation}
This is not satisfied for generic solutions. Hence in a spacetime
for closed hypersurfaces, the theory is generically linearization
unstable.

\section{Conclusions}

We have constructed the gauge-invariant conserved electric and magnetic
charges in Yang-Mills theory in a dynamical curved background generalizing
the flat spacetime construction of Abbott-Deser \citep{Abbott-Deser-nonabelian}.
Electric charges arise from the field equations, while the magnetic
charges arise from the Bianchi identity. The crucial ingredient is
the symmetry of the background gauge field that solves the curved
space Yang-Mills equation. For the gravity part one can take any geometric
theory of gravity based on the Riemannian geometry, but to be concrete
we chose the cosmological General Relativity. To be able to define
the electric and magnetic charges, besides the mentioned symmetry
of the background gauge field, as defined by $\delta_{\xi}A_{\mu}=\bar{D}_{\mu}\xi=0$,
one also needs a time-like Killing vector for the spacetime which
we assumed. Our results in curved spacetime reduces to the flat spacetime
expressions in the correct limit.

We have also studied the linearization instability issue in the Gravity-Yang-Mills
theory and established a second order integral constraint that must
be satisfied by any solution of linearized Yang-Mills theory in a
spacetime with closed (compact without boundary) spatial hypersurfaces.
We have not discussed the linearization instability in the gravity
sector as it was recently done in \cite{altasuzunmakale} and described
in great detail in the thesis \cite{emeltez}.

\section{Acknowledgments}

The works of E.A. and E.K. are partially supported by the TUBITAK
Grant No. 120F253. The work of E.K. is partially supported by the
TUBITAK Grant No. 119F241.

\section{Appendix a: First and second order Expansions of the field equations}

Here, we consider the expansion of the Yang-Mills fields and equations
about the background quantities, the background metric and background
gauge field, up to the cubic terms. While the first order terms will
be used to construct the conserved charges, the quadratic terms will
give us the integral constraint on solutions of the linearized equations.

Let us start with the gauge field and assume that it can be expanded
about the background field $\bar{\hat{A}}_{\mu}$ up to the third
order terms as 
\begin{equation}
\hat{A}_{\mu}=\bar{\hat{A}}_{\mu}+\lambda\hat{a}_{\mu}+\frac{\lambda^{2}}{2}\hat{b}_{\mu}.\label{expansionofgaugefiedl-1}
\end{equation}
Here $\lambda$ denotes the expansion parameter, $\hat{a}_{\mu}$
and $\hat{b}_{\mu}$ are the first and the second order expansions
respectively. The background gauge field $\bar{\hat{A}}_{\mu}$ satisfies
the background field equations without a source 
\begin{equation}
\bar{D}_{\mu}\bar{\hat{F}}^{\mu\nu}=\bar{\nabla}_{\mu}\bar{\hat{F}}^{\mu\nu}+[\bar{\hat{A}}_{\mu},\bar{\hat{F}}^{\mu\nu}]=0.\label{background field equations-1}
\end{equation}
We express the expansion of a generic tensor field $T$ about its
background value $\bar{T}$ as 
\begin{equation}
T=\bar{T}+(T)^{(1)}+(T)^{(2)}+...,
\end{equation}
where $(T)^{(1)}$ denotes the linearized $T$ tensor and $(T)^{(2)}$
denotes the second order expansion of it. For example, explicitly
the field strength is expanded as 
\begin{equation}
\hat{F}_{\mu\nu}=\bar{\hat{F}}_{\mu\nu}+(\hat{F}_{\mu\nu})^{(1)}+(\hat{F}_{\mu\nu})^{(2)}+...
\end{equation}
up to the third order. Assuming a Riemann connection, we have 
\begin{equation}
\hat{F}_{\mu\nu}=\partial_{\mu}\hat{A}_{\nu}-\partial_{\nu}\hat{A}_{\mu}+[\hat{A}_{\mu},\hat{A}_{\nu}].
\end{equation}
The decomposition of the field strength at first order reads 
\begin{equation}
(\hat{F}_{\mu\nu})^{(1)}=\lambda(\bar{D}_{\mu}\hat{a}_{\nu}-\bar{D}_{\nu}\hat{a}_{\mu}),
\end{equation}
and at second order one arrives at 
\begin{equation}
(\hat{F}_{\mu\nu})^{(2)}=\frac{\lambda^{2}}{2}\left(\bar{D}_{\mu}\hat{b}_{\nu}-\bar{D}_{\nu}\hat{b}_{\mu}+2[\hat{a}_{\mu},\hat{a}_{\nu}]\right).
\end{equation}
Now we can compute the expansion of $\hat{F}^{\mu\nu}$. For this
purpose, we use perturbation of the spacetime metric about a background
metric $\bar{g}_{\mu\nu}$ 
\begin{equation}
g_{\mu\nu}=\bar{g}_{\mu\nu}+\tau h_{\mu\nu}+\frac{\tau^{2}}{2}k_{\mu\nu},
\end{equation}
and its inverse 
\begin{equation}
g^{\mu\nu}=\bar{g}^{\mu\nu}-\tau h^{\mu\nu}+\frac{\tau^{2}}{2}\left(2h^{\mu\sigma}h_{\sigma}^{\nu}-k^{\mu\nu}\right).
\end{equation}
The field strength with upper indices, $\hat{F}^{\mu\nu}=g^{\mu\sigma}g^{\nu\rho}\hat{F}_{\sigma\rho}$,
at the first order yields 
\begin{equation}
(\hat{F}^{\mu\nu})^{(1)}=\lambda(\bar{D}^{\mu}\hat{a}^{\nu}-\bar{D}^{\nu}\hat{a}^{\mu})+\tau(\bar{\hat{F}}^{\sigma\mu}h_{\sigma}^{\nu}-\bar{\hat{F}}^{\sigma\nu}h_{\sigma}^{\mu}),
\end{equation}
which at second order reads 
\begin{eqnarray}
 &  & (\hat{F}^{\mu\nu})^{(2)}=\frac{\lambda^{2}}{2}\left(\bar{D}^{\mu}\hat{b}^{\nu}-\bar{D}^{\nu}\hat{b}^{\mu}+2\left[\hat{a}^{\mu},\hat{a}^{\nu}\right]\right)+\tau\lambda\left(h^{\nu\sigma}(\bar{D}_{\sigma}\hat{a}^{\mu}-\bar{D}^{\mu}\hat{a}_{\sigma})+h^{\mu\sigma}(\bar{D}^{\nu}\hat{a}_{\sigma}-\bar{D}_{\sigma}\hat{a}^{\nu})\right)\nonumber \\
 &  & ~~~~~~~~~~~+\frac{\tau^{2}}{2}\left(\bar{\hat{F}}^{\mu\sigma}(2h^{\nu\lambda}h_{\lambda\sigma}-k_{\sigma}^{\nu})-\bar{\hat{F}}^{\nu\sigma}(2h^{\mu\lambda}h_{\lambda\sigma}-k_{\sigma}^{\mu})+2\bar{\hat{F}}_{\sigma\rho}h^{\mu\sigma}h^{\nu\rho}\right).
\end{eqnarray}
Now we can expand $D_{\mu}\hat{F}^{\mu\nu}$. Using 
\begin{equation}
D_{\mu}\hat{F}^{\mu\nu}=\partial_{\mu}\hat{F}^{\mu\nu}+\Gamma_{\mu\sigma}^{\mu}\hat{F}^{\sigma\nu}+[\hat{A}_{\mu},\hat{F}^{\mu\nu}]
\end{equation}
together with the previous expressions one obtains 
\begin{equation}
(D_{\mu}\hat{F}^{\mu\nu})^{(1)}=\bar{D}_{\mu}\left(\lambda(\bar{D}^{\mu}\hat{a}^{\nu}-\bar{D}^{\nu}\hat{a}^{\mu})+\tau(\bar{\hat{F}}^{\sigma\mu}h_{\sigma}^{\nu}-\bar{\hat{F}}^{\sigma\nu}h_{\sigma}^{\mu})+\frac{\tau}{2}\bar{\hat{F}}^{\mu\nu}h\right)+\lambda[\hat{a}_{\mu},\bar{\hat{F}}^{\mu\nu}],\label{D_muF^munu-firstorder-1}
\end{equation}
where $h=\bar{g}^{\mu\nu}h_{\mu\nu}$. Similarly, the second order
expansion gives us the following 
\begin{eqnarray}
 &  & (D_{\mu}\hat{F}^{\mu\nu})^{(2)}=\bar{D}_{\mu}\left((\hat{F}^{\mu\nu})^{(2)}+\frac{\tau}{2}(\hat{F}^{\mu\nu})^{(1)}h+\frac{\tau^{2}}{4}\bar{\hat{F}}^{\mu\nu}(k-h_{\rho\sigma}h^{\rho\sigma})\right)\\
 &  & ~~~~~~~~~~~~~~~~~~-\frac{\tau}{2}h\bar{D}_{\mu}(\hat{F}^{\mu\nu})^{(1)}+\lambda[\hat{a}_{\mu},(\hat{F}^{\mu\nu})^{(1)}]+\frac{\lambda^{2}}{2}[\hat{b}_{\mu},\bar{\hat{F}}^{\mu\nu}],\nonumber 
\end{eqnarray}
with $k=\bar{g}^{\mu\nu}k_{\mu\nu}$. In order to construct the conserved
charges of the theory, we will not use the explicit form of the second
order expansion. But this result will become important in linearization
instability discussion. The field equations are expanded as 
\begin{equation}
D_{\mu}\hat{F}^{\mu\nu}=\bar{D}_{\mu}\bar{\hat{F}}^{\mu\nu}+(D_{\mu}\hat{F}^{\mu\nu})^{(1)}+(D_{\mu}\hat{F}^{\mu\nu})^{(2)}+...=\hat{J}^{\nu}
\end{equation}
where $\bar{D}_{\mu}\bar{\hat{F}}^{\mu\nu}=0$ by assumption. We put
all the higher order terms to the right hand side of the equation
and define a new current 
\begin{equation}
\hat{{\mathcal{J}}}^{\nu}:=\hat{J}^{\nu}-(D_{\mu}\hat{F}^{\mu\nu})^{(2)}-...\thinspace.
\end{equation}
Then we express the linearized field equations as 
\begin{equation}
(D_{\mu}\hat{F}^{\mu\nu})^{(1)}=\hat{{\mathcal{J}}}^{\nu}.
\end{equation}
Substituting (\ref{D_muF^munu-firstorder-1}) in the last equation,
one finds 
\begin{equation}
\bar{D}_{\mu}\left(\lambda(\bar{D}^{\mu}\hat{a}^{\nu}-\bar{D}^{\nu}\hat{a}^{\mu})+\tau(\bar{\hat{F}}^{\sigma\mu}h_{\sigma}^{\nu}-\bar{\hat{F}}^{\sigma\nu}h_{\sigma}^{\mu})+\frac{\tau}{2}\bar{\hat{F}}^{\mu\nu}h\right)+\lambda[\hat{a}_{\mu},\bar{\hat{F}}^{\mu\nu}]=\hat{{\mathcal{J}}}^{\nu}.\label{linearizedfield equations-1}
\end{equation}
Using the last equation one can prove the conservation of the new
current $\hat{{\mathcal{J}}}^{\nu}$.

\section{APPENDIX B: CONSERVATION OF THE NEW CURRENT}

For the consistency of the construction, the new current have to be
conserved. To prove the conservation let us consider an antisymmetric
rank two tensor, say $X^{\mu\nu}$. We first calculate the commutator
$\left[\bar{D}_{\nu},\bar{D}_{\mu}\right]X^{\mu\nu}$ to make the
construction easier. Explicitly we write 
\begin{equation}
[\bar{D}_{\nu},\bar{D}_{\mu}]X^{\mu\nu}=\bar{D}_{\nu}\bar{D}_{\mu}X^{\mu\nu}-\bar{D}_{\mu}\bar{D}_{\nu}X^{\mu\nu},
\end{equation}
which yields

\begin{equation}
[\bar{D}_{\nu},\bar{D}_{\mu}]X^{\mu\nu}=[\bar{\nabla}_{\nu},\bar{\nabla}_{\mu}]X^{\mu\nu}+[\bar{\nabla}_{\nu}\bar{\hat{A}}_{\mu}-\bar{\nabla}_{\mu}\bar{\hat{A}}_{\nu},X^{\mu\nu}]+[\bar{\hat{A}}_{\nu},[\bar{\hat{A}}_{\mu},X^{\mu\nu}]]-[\bar{\hat{A}}_{\mu},[\bar{\hat{A}}_{\nu},X^{\mu\nu}]],\label{commutator-1}
\end{equation}
where $[\bar{\nabla}_{\nu},\bar{\nabla}_{\mu}]X^{\mu\nu}=0$. Using
the Jacobi identity, the last two terms in the last equation yields
\begin{equation}
[\bar{\hat{A}}_{\nu},[\bar{\hat{A}}_{\mu},X^{\mu\nu}]]-[\bar{\hat{A}}_{\mu},[\bar{\hat{A}}_{\nu},X^{\mu\nu}]]=-[X^{\mu\nu},[\bar{\hat{A}}_{\nu},\bar{\hat{A}}_{\mu}]].
\end{equation}
Then equation (\ref{commutator-1}) reduces to 
\begin{equation}
[\bar{D}_{\nu},\bar{D}_{\mu}]X^{\mu\nu}=[\bar{\nabla}_{\nu}\bar{\hat{A}}_{\mu}-\bar{\nabla}_{\mu}\bar{\hat{A}}_{\nu},X^{\mu\nu}]-[X^{\mu\nu},[\bar{\hat{A}}_{\nu},\bar{\hat{A}}_{\mu}]].
\end{equation}
Since $\bar{\nabla}_{\nu}\bar{\hat{A}}_{\mu}-\bar{\nabla}_{\mu}\bar{\hat{A}}_{\nu}=\bar{\hat{F}}_{\nu\mu}-[\bar{\hat{A}}_{\nu},\bar{\hat{A}}_{\mu}]$,
one can re-express the commutator as 
\begin{equation}
[\bar{D}_{\nu},\bar{D}_{\mu}]X^{\mu\nu}=[\bar{\hat{F}}_{\nu\mu},X^{\mu\nu}].\label{identity1-1}
\end{equation}
Due to antisymmetry of the tensor field $X^{\mu\nu}$, the last expression
also yields the following identity 
\begin{equation}
\bar{D}_{\nu}\bar{D}_{\mu}X^{\mu\nu}=\frac{1}{2}[\bar{\hat{F}}_{\nu\mu},X^{\mu\nu}].\label{identity2-1}
\end{equation}
For the special case $X^{\mu\nu}=\hat{F}^{\mu\nu}$, one has 
\begin{equation}
\bar{D}_{\nu}\bar{D}_{\mu}\hat{F}^{\mu\nu}=\frac{1}{2}[\bar{\hat{F}}_{\nu\mu},\hat{F}^{\mu\nu}]=0.\label{identity3-1}
\end{equation}
Note that $\bar{D}_{\nu}\hat{{\mathcal{J}}}^{\nu}$ includes these
type of terms and the above identities will be useful when we prove
the conservation of the new current. From equation (\ref{linearizedfield equations-1}),
we write 
\begin{equation}
\bar{D}_{\nu}\hat{{\mathcal{J}}}^{\nu}=\bar{D}_{\nu}\bar{D}_{\mu}\left(\lambda(\bar{D}^{\mu}\hat{a}^{\nu}-\bar{D}^{\nu}\hat{a}^{\mu})+\tau(\bar{\hat{F}}^{\sigma\mu}h_{\sigma}^{\nu}-\bar{\hat{F}}^{\sigma\nu}h_{\sigma}^{\mu})+\frac{\tau}{2}\bar{\hat{F}}^{\mu\nu}h\right)+\lambda\bar{D}_{\nu}[\hat{a}_{\mu},\bar{\hat{F}}^{\mu\nu}].
\end{equation}
Using the identity (\ref{identity2-1}) it can be rewritten as 
\begin{equation}
\bar{D}_{\nu}\hat{{\mathcal{J}}}^{\nu}=\frac{\lambda}{2}[\bar{D}^{\mu}\hat{a^{\nu}}-\bar{D}^{\nu}\hat{a}^{\mu},\bar{\hat{F}}_{\mu\nu}]+\frac{\tau}{2}[\bar{\hat{F}}^{\sigma\mu}h_{\sigma}^{\nu}-\bar{\hat{F}}^{\sigma\nu}h_{\sigma}^{\mu},\bar{\hat{F}}_{\mu\nu}]+\frac{\tau}{4}h[\bar{\hat{F}}^{\mu\nu},\bar{\hat{F}}_{\mu\nu}]+\lambda[\bar{D}_{\nu}\hat{a}_{\mu},\bar{\hat{F}}^{\mu\nu}],
\end{equation}
and then it becomes 
\begin{equation}
\bar{D}_{\nu}\hat{{\mathcal{J}}}^{\nu}=\lambda[\bar{D}^{\mu}\hat{a}^{\nu},\bar{\hat{F}}_{\mu\nu}]+\tau h_{\sigma}^{\nu}[\bar{\hat{F}}^{\sigma\mu},\bar{\hat{F}}_{\mu\nu}]+\frac{\tau}{4}h[\bar{\hat{F}}^{\mu\nu},\bar{\hat{F}}_{\mu\nu}]+\lambda[\bar{D}_{\nu}\hat{a}_{\mu},\bar{\hat{F}}^{\mu\nu}].
\end{equation}
The first and the last term on the right vanish from the antisymmetry
of the indices. Also we have proved the vanishing of the third term
in equation (\ref{identity3-1}). There remains the second term only
\begin{equation}
\bar{D}_{\nu}\hat{{\mathcal{J}}}^{\nu}=\tau h_{\sigma}^{\nu}[\bar{\hat{F}}^{\sigma\mu},\bar{\hat{F}}_{\mu\nu}].
\end{equation}
Renaming the indices $\nu$ and $\sigma$, vanishing of this term
is obvious. So, one ends up with $\bar{D}_{\nu}\hat{{\mathcal{J}}}^{\nu}=0$,
which is the expected result.

\section{APPENDIX C: DEFINITION OF THE CONSERVED CHARGES}

Using the expressions $\bar{D}_{\nu}\hat{{\mathcal{J}}}^{\nu}=0$
and $\bar{D}_{\nu}\bar{\hat{\xi}}^{s}=0$, we can write 
\begin{equation}
\bar{D}_{\nu}(\bar{\hat{\xi}}^{s}\hat{{\mathcal{J}}}^{\nu})=0=\bar{\nabla}_{\nu}(\bar{\hat{\xi}}^{s}\hat{{\mathcal{J}}}^{\nu})+[\bar{\hat{A}}_{\nu},\bar{\hat{\xi}}^{s}\hat{{\mathcal{J}}}^{\nu}].
\end{equation}
But we need a quantity which is conserved in the ordinary sense instead
of the covariant conservation. Following the flat spacetime case we
write 
\begin{equation}
\bar{D}_{\nu}\text{Tr}(\bar{\hat{\xi}}^{s}\hat{{\mathcal{J}}}^{\nu})=\bar{\nabla}_{\nu}\text{Tr}(\bar{\hat{\xi}}^{s}\hat{{\mathcal{J}}}^{\nu})+[\bar{\hat{A}}_{\nu},\text{Tr}(\bar{\hat{\xi}}^{s}\hat{{\mathcal{J}}}^{\nu})]=0,
\end{equation}
where $[\bar{\hat{A}}_{\nu},\text{Tr}(\bar{\hat{\xi}}^{s}\hat{{\mathcal{J}}}^{\nu})]=0$
and so we have 
\begin{equation}
\bar{D}_{\nu}\text{Tr}(\bar{\hat{\xi}}^{s}\hat{{\mathcal{J}}}^{\nu})=\bar{\nabla}_{\nu}\text{Tr}(\bar{\hat{\xi}}^{s}\hat{{\mathcal{J}}}^{\nu})=0.
\end{equation}
Multiplying with $\sqrt{-\bar{g}}$ and using $\sqrt{-\bar{g}}\bar{\nabla}_{\nu}X^{\nu}=\partial_{\nu}\left(\sqrt{-\bar{g}}X^{\nu}\right)$
we express 
\begin{equation}
\sqrt{-\bar{g}}\bar{D}_{\nu}\text{Tr}(\bar{\hat{\xi}}^{s}\hat{{\mathcal{J}}}^{\nu})=\sqrt{-\bar{g}}\bar{\nabla}_{\nu}\text{Tr}(\bar{\hat{\xi}}^{s}\hat{{\mathcal{J}}}^{\nu})=\partial_{\nu}\left(\sqrt{-\bar{g}}\text{Tr}(\bar{\hat{\xi}}^{s}\hat{{\mathcal{J}}}^{\nu})\right),
\end{equation}
from which we can define the total charges as 
\begin{equation}
Q^{s}:=\frac{1}{4\pi}\intop d^{4}x~\partial_{0}\left(\sqrt{-\bar{g}}\text{Tr}(\bar{\hat{\xi}}^{s}\hat{{\mathcal{J}}}^{0})\right).
\end{equation}
Using the Stokes theorem this can be written as 
\begin{equation}
Q^{s}:=\frac{1}{4\pi}\intop d^{3}x~\sqrt{\bar{\gamma}}\text{Tr}(\bar{\hat{\xi}}^{s}\hat{{\mathcal{J}}}^{0}).\label{conservedcharges1-1}
\end{equation}
Note that $\bar{\gamma}$ denotes the induced metric on the hypersurface.
Using the explicit form of the linearized field equations $\hat{{\mathcal{J}}}^{0}$
reads 
\begin{equation}
\hat{{\mathcal{J}}}^{0}=\bar{D}_{i}\left(\lambda(\bar{D}^{i}\hat{a}^{0}-\bar{D}^{0}\hat{a}^{i})+\tau(\bar{\hat{F}}^{0i}h_{0}^{0}+\bar{\hat{F}}^{ki}h_{k}^{0}+\bar{\hat{F}}^{0k}h_{k}^{i}+\frac{h}{2}\bar{\hat{F}}^{i0})\right)+\lambda[\hat{a}_{i},\bar{\hat{F}}^{i0}].\label{current_0-1}
\end{equation}
We have 
\begin{equation}
\text{Tr}\left(\bar{\hat{\xi}}^{s}[\hat{a}_{i},\bar{\hat{F}}^{i0}]\right)=\text{Tr}\left([\bar{\hat{\xi}}^{s},\bar{\hat{F}}^{i0}]\hat{a}_{i}\right)=0,
\end{equation}
where the first equality comes form the cyclic property of trace and
the second one is obtained from $[\bar{D}_{\mu},\bar{D}_{\nu}]\bar{\hat{\xi}}^{s}=0$.
Then inserting (\ref{current_0-1}) in equation (\ref{conservedcharges1-1}),
the conserved charges can be written as 
\begin{equation}
Q^{s}:=\frac{1}{4\pi}\intop d^{3}x~\sqrt{\bar{\gamma}}\text{Tr}\bar{D}_{i}\left(\bar{\hat{\xi}}^{s}\Bigl(\lambda(\bar{D}^{i}\hat{a}^{0}-\bar{D}^{0}\hat{a}^{i})+\tau(\bar{\hat{F}}^{0i}h_{0}^{0}+\bar{\hat{F}}^{ki}h_{k}^{0}+\bar{\hat{F}}^{0k}h_{k}^{i}+\frac{h}{2}\bar{\hat{F}}^{i0})\Bigr)\right).
\end{equation}
To be able to use the Stokes theorem again, we need to convert the
background gauge covariant derivative to the tensorial covariant derivative.
The gauge covariant derivative and trace commute with each other.
So we can express 
\begin{eqnarray}
 &  & 4\pi Q^{s}:=\frac{}{}\intop d^{3}x~\sqrt{\bar{\gamma}}\bar{\nabla}_{i}\text{Tr}\left(\bar{\hat{\xi}}^{s}\Bigl(\lambda(\bar{D}^{i}\hat{a}^{0}-\bar{D}^{0}\hat{a}^{i})+\tau(\bar{\hat{F}}^{0i}h_{0}^{0}+\bar{\hat{F}}^{ki}h_{k}^{0}+\bar{\hat{F}}^{0k}h_{k}^{i}+\frac{h}{2}\bar{\hat{F}}^{i0})\Bigr)\right)\nonumber \\
 &  & +\intop d^{3}x~\sqrt{\bar{\gamma}}\left[\bar{\hat{A}}_{i},\text{Tr}\left(\bar{\hat{\xi}}^{s}\Bigl(\lambda(\bar{D}^{i}\hat{a}^{0}-\bar{D}^{0}\hat{a}^{i})+\tau(\bar{\hat{F}}^{0i}h_{0}^{0}+\bar{\hat{F}}^{ki}h_{k}^{0}+\bar{\hat{F}}^{0k}h_{k}^{i}+\frac{h}{2}\bar{\hat{F}}^{i0})\Bigr)\right)\right],
\end{eqnarray}
where the terms in the second line of the last equation vanish automatically.
Then we arrive at 
\begin{equation}
Q^{s}:=\frac{1}{4\pi}\intop d^{3}x~\partial_{i}\left\{ \sqrt{\bar{\gamma}}\text{Tr}\left(\bar{\hat{\xi}}^{s}\Bigl(\lambda(\bar{D}^{i}\hat{a}^{0}-\bar{D}^{0}\hat{a}^{i})+\tau(\bar{\hat{F}}^{0i}h_{0}^{0}+\bar{\hat{F}}^{ki}h_{k}^{0}+\bar{\hat{F}}^{0k}h_{k}^{i}+\frac{h}{2}\bar{\hat{F}}^{i0})\Bigr)\right)\right\} .
\end{equation}
After applying the Stokes theorem one more time the last equation
yields the following expression for the conserved charges 
\begin{equation}
Q^{s}:=\frac{1}{4\pi}\intop d^{2}x~\sqrt{\bar{\beta}}\bar{\sigma}_{i}\text{Tr}\left(\bar{\hat{\xi}}^{s}\Bigl(\lambda(\bar{D}^{i}\hat{a}^{0}-\bar{D}^{0}\hat{a}^{i})+\tau(\bar{\hat{F}}^{0i}h_{0}^{0}+\bar{\hat{F}}^{ki}h_{k}^{0}+\bar{\hat{F}}^{0k}h_{k}^{i}+\frac{h}{2}\bar{\hat{F}}^{i0})\Bigr)\right).
\end{equation}
Here $\bar{\beta}$ denotes the two dimensional induced metric on
the boundary of the hypersurface and $\bar{\sigma}_{i}$ is its unit
one form.


\begin{thebibliography}{10}
\bibitem{Abbott-Deser-nonabelian}L.F. Abbott, S. Deser, Charge Definition
in Nonabelian Gauge Theories, Phys. Lett. B \textbf{116}, 259-263
(1982).

\bibitem{Einstein-Pauli}A. Einstein and W. Pauli, On the Non-Existence
of Regular Stationary Solutions of Relativistic Field Equations, Annals
of Mathematics Second Series,\textbf{ 44}, 2 (1943).

\bibitem{Deser1976}S. Deser, Absence of Static Solutions in Source-Free
Yang-Mills Theory, Phys. Lett. B \textbf{64} 463-464 (1976).

\bibitem{Coleman}S. Coleman, There are no classical glueballs, Comm.
Math. Phys. \textbf{55} (2), 113-116 (1977).

\bibitem{Bartnik-Mckinnon}R. Bartnik, J. Mckinnon, Particle - Like
Solutions of the Einstein Yang-Mills Equations, Phys. Rev. Lett. \textbf{61},
141-144 (1988).

\bibitem{Hosotani}J. Bjoraker, Y. Hosotani, Stable monopole and dyon
solutions in the Einstein-Yang-Mills theory in asymptotically anti-de
Sitter Space, Phys.Rev.Lett. \textbf{84}, 1853 (2000).

\bibitem{Edery-Nakayama} A. Edery, Y. Nakayama, Gravitating magnetic
monopole via the spontaneous symmetry breaking of pure $R^{2}$ gravity,
Phys. Rev. D \textbf{98}, 064011 (2018).

\bibitem{ADM}R. Arnowitt, S. Deser and C. W. Misner, Canonical variables
for general relativity, Phys. Rev.\textbf{ 117}, 1595 (1960); The
dynamics of general relativity, Gen. Rel. Grav.\textbf{ 40}, 1997
(2008).

\bibitem{AD}L. F. Abbott and S. Deser, Stability of gravity with
a cosmological constant, Nucl. Phys. B \textbf{195}, 76 (1982).

\bibitem{DT} S. Deser and B. Tekin, Energy in generic higher curvature
gravity theories, Phys. Rev. D \textbf{67}, 084009 (2003); Gravitational
energy in quadratic curvature gravities, Phys. Rev. Lett. \textbf{89},
101101 (2002).

\bibitem{Fischer-Marsden}A. E. Fischer, J. E. Marsden, Linearization
stability of the Einstein equations, Bulletin of the American Mathematical
Society \textbf{79}, 997-1003 (1973).

\bibitem{Deser-Brill} S. Deser S, D. Brill, Instability of closed
spaces in general relativity, Communications in Mathematical Physics
\textbf{32}, 291--304 (1973).

\bibitem{Deser-Bruhat}S. Deser, Y. Choquet-Bruhat, On the stability
of flat space, Annals of Physics \textbf{81}, 165-178 (1973).

\bibitem{Moncrief} V. Moncrief, Spacetime symmetries and linearization
stability of the Einstein equations I, Journal of Mathematical Physics
\textbf{16}, 493-498 (1975).

\bibitem{Arms-Marsden} J. M. Arms, J. E. Marsden, The absence of
Killing fields is necessary for linearization stability of Einstein's
equations,\enskip{}Indiana University Mathematics Journal \textbf{28},
119-125 (1979).

\bibitem{Fischer-Marsden-Moncrief}A. E. Fischer, J. E. Marsden, V.
Moncrief, The structure of the space of solutions of Einstein's equations.
I. One Killing field, Annales de l'I.H.P. Physique theorique \textbf{33},
147-194 (1980).

\bibitem{Marsden} J. E. Marsden, Lectures on geometric methods in
mathematical physics, CBMS-NSF Regional Conference Series in Applied
Mathematics, SIAM, Philadelphia, 37 (1981).

\bibitem{Girbau-Bruna} J. Girbau, L. Bruna, Stability by linearization
of Einstein's field equation, Springer (2010).

\bibitem{altasuzunmakale} E. Altas, B. Tekin, Linearization instability
for generic gravity in AdS spacetime, Physical Review D \textbf{97},
024028 (2018).

\bibitem{emelchiral} E. Altas, B. Tekin, Linearization instability
of chiral gravity, Physical Review D \textbf{97}, 124068 (2018).

\bibitem{emeltez}E. Altas, Linearization Instability in Gravity Theories,
PhD Thesis, Middle East Tech. U., Ankara (2018), arXiv:1808.04722
{[}hep-th{]}.

\bibitem{Weinberg}S. Weinberg, The Quantum Theory of Fields, Volume
2, Cambridge University Press, (1996).

\bibitem{Taub} A. H. Taub, Variational principles in general relativity,
Lectures at the Centro Internazionale Matematico Estiud, 206-300 (1970).

\bibitem{emelson} E. Altas, B. Tekin, Second order perturbation theory
in general relativity: Taub charges as integral constraints, Physical
Review D \textbf{99}, 104078 (2019). 
\end{thebibliography}
\end{document}